\documentclass[11pt]{article}
\usepackage{amsthm,amsmath,amssymb,bm,multirow,color,natbib}
\usepackage{fancyhdr}
\usepackage[hypertexnames=false,bookmarksnumbered]{hyperref}
\hypersetup{colorlinks=true,citecolor=blue}
\usepackage{graphicx}
\usepackage{algorithm}

\newcommand{\cip}{\overset{P}{\longrightarrow}}
\newcommand{\cid}{\overset{D}{\longrightarrow}}
\newcommand{\Real}{\mathbb R}
\newcommand{\bX}{{\mathbf X}}

\newcommand{\bY}{{\mathbf Y}}
\newcommand{\by}{{\mathbf y}}

\newcommand{\bu}{{\mathbf u}}
\newcommand{\bU}{{\mathbf U}}

\newcommand{\bG}{{\mathbf G}}

\newcommand{\bi}{{\mathbf i}}

\newcommand{\defeq}{\overset{\text{def}}{=}}

\newcommand{\bbeta}{{\mbox{\boldmath $\beta$}}}

\newcommand{\btheta}{{\mbox{\boldmath $\theta$}}}
\newcommand{\bTheta}{{\mbox{\boldmath $\Theta$}}}
\newcommand{\Rtheta}{{\mbox{\boldmath $\vartheta$}}}
\newcommand{\aname}{{Method G}}

\theoremstyle{plain}
\newtheorem{prop}{\emph{Proposition}}

\theoremstyle{definition}

\theoremstyle{remark}

\newcommand{\ignore}[1]{}

\begin{document}
\title{\vspace*{-0cm}\aname: Uncertainty Quantification for Distributed Data Problems using Generalized Fiducial Inference}

\author{
Randy C. S. Lai\thanks{Department of Mathematics \& Statistics, University of Maine, 5752 Neville Hall, Orono, ME 04469, USA.  Email: {\ttfamily chushing.lai@maine.edu}}
\and
Jan Hannig\thanks{Corresponding author.  Department of Statistics and Operations Research, University of North Carolina at Chapel Hill, Chapel Hill, NC 27599-3260, USA.  Email: {\ttfamily jan.hannig@unc.edu}}
\and
Thomas C. M. Lee\thanks{Department of Statistics, University of California at Davis, 4118 Mathematical Sciences Building, One Shields Avenue, Davis, CA 95616, USA. Email: {\ttfamily tcmlee@ucdavis.edu}} \\
}

\date{March 8, 2018}

\maketitle

\begin{abstract}
It is not unusual for a data analyst to encounter data sets distributed across several computers.  This can happen for reasons such as privacy concerns, efficiency of likelihood evaluations, or just the sheer size of the whole data set. This presents new challenges to statisticians as even computing simple summary statistics such as the median becomes computationally challenging.  Furthermore, if other advanced statistical methods are desired, novel computational strategies are needed.  In this paper we propose a new approach for distributed analysis of massive data that is suitable for generalized fiducial inference and is based on a careful implementation of a ``divide and conquer'' strategy combined with importance sampling.  The proposed approach requires only small amount of communication between nodes, and is shown to be asymptotically equivalent to using the whole data set.  Unlike most existing methods, the proposed approach produces uncertainty measures (such as confidence intervals) in addition to point estimates for parameters of interest. The proposed approach is also applied to the analysis of a large set of solar images.

Keywards: divide and conquer, importance sampling, massive data, parallel processing, solar imaging
\end{abstract}

\section{Introduction}
The increased availability of cloud computing brings new challenges to practical data analysis.  For example, advances in modern science and business allow the collection of massive data sets. An example is high throughput sequencing in genetics that is capable of producing terabytes of information in a single experiment. Even if the data set itself is not massive, there are other reasons that it needs to be analyzed in a distributive manner. For example privacy concerns might require data sets to stay within the country or company of origin and share summary information only. Similarly computational efficiency of MCMC algorithms sometimes deteriorates with the sample size so one might want to run multiple MCMC chains on different portions of the data.

This presents new challenges to statisticians as even computing simple summary statistics such as the median of such a data set becomes computationally challenging.  If other advanced statistical methods are required for analyzing such data sets, novel computational strategies are needed.  An appealing approach to analyzing a massive data set is the so-called ``divide and conquer'' strategy. That is, the data set is first divided into manageable subsets, then each subset is analyzed separately, often on a parallel computer, and finally the results of the analyses are combined.

In order to efficiently combine the results from the various subgroups, one needs to account for the uncertainties in the estimates based on each of the subsets. Among the frequentist proposals, \cite{KleinerEtAl2011} propose a parallelized version of bootstrap, \cite{ChenXie2014} propose the use of confidence distributions, and \cite{battey2015distributed} perform distributed testing.  In the Bayesian literature many recent algorithms propose using the embarrassingly parallel approach with various modifications to assess with combination of the results afterward.  \cite{NeiswangerWangXing2014} and \cite{LeisenCraiuCesarin2016} propose using normal approximation to the posterior while \cite{WangGuoHellerDunson2015} and \cite{srivastava2015wasp} advocate for a more advanced combination based on the Wasserstein distance.

In this paper we propose a new approach that is suitable for generalized fiducial inference (GFI), which have proven to provide a distribution on the parameter space with good inferential properties without the need for a subjective prior specification \citep{hannig2015review}. Our parallel algorithm uses minimal amount of information swapping between workers to improve efficiency of the algorithm while maintaining the ability to run different MCMC algorithms on each worker. We then use a careful importance sampling scheme to combine the results from various workers. Our method produces uncertainty measures (such as confidence intervals) as well as point estimates for the parameters of interest.  We prove consistency and asymptotic normality of the approximation scheme and provide numerical comparisons showing good performance of our algorithm. While the proposed method has been designed for GFI it is also applicable for Bayes posteriors.  We call our proposal {\em \aname}.

The rest of this article is organized as follows.  First, some background material for GFI is provided in Section~\ref{sec:background}. Then the proposed methodology is developed in Section~\ref{sec:massive}, which include theoretical backup and a practical algorithm. The finite sample performance of the proposed methodology is illustrated via numerical experiments in Section~\ref{sec:simulation} and real data application in Section~\ref{sec:solar}. Lastly, concluding remarks are offered in Section~\ref{sec:conclude} while technical details are deferred to the appendix.

\section{Background of Generalized Fiducial Inference}
\label{sec:background}
\cite{Fisher1930} introduced fiducial inference in the hope to define a distribution on the parameter space when the Bayesian approach cannot be applied due to the lack of a suitable prior.  Unfortunately his fiducial proposal carried some defects and hence was not welcomed by the statistics community. Generalized fiducial inference (GFI) is an improved version of Fisher's idea that rectifies these defects. See \cite{hannig2015review} for an up-to-date review of GFI.

Suppose we have $\bY=\{Y_1,\ldots,Y_n\}$ i.i.d continuous random variables from some distribution $F
(y;\btheta)$ with an unknown $p$-variate parameter $\btheta$ and parameter space $\bTheta$; i.e.,
$\btheta \in \bTheta \subset \mathbb \Re^p$.  Denote the corresponding density function as $f(y;\btheta)$.
It is further supposed that the observation vector $
\bY$ could be written as a mapping from a pivotal random vector $\bU = \{U_1,\ldots,U_n\}$ such
that
\begin{equation}
\bY = \bG(\btheta, \bU).
\end{equation}

Inverting this data generating equation provides us with a generalized fiducial density $r(\btheta)$: a distribution
on the parameter space obtained without the need to define a prior distribution.
\cite{hannig2015review} showed that the generalized fiducial density $r(\btheta; \by)$ of $\btheta$ for a fixed observed data $\bY = \by$ is
\begin{align} \label{eq:overall.fiducial}
r(\btheta ; \by)  = \frac{ f(\by ; \btheta) J(\by, \btheta)}{\int f(\by ; \btheta') J(\by, \btheta')
d\btheta'} \defeq \frac{1}{c(\by)} f(\by ; \btheta) J(\by, \btheta),
\end{align}
where
\begin{equation}\label{eq:RecommendedJacobian}
 J(\by,\btheta)= D\left( \frac{d}{ d\btheta} \bG(\bu, \btheta) \Big | _{u = \bG^{-1}(\by,
 \btheta)}\right).
\end{equation}

The $D$ function has two canonical forms derived in \citep{hannig2015review}. The form of $D$ depends on how we define neighborhoods of the observed data $y$. The first uses neighborhoods specified by the $L_{\infty}$ norm (corresponding to observing discretized data) and the
resulting $D$ is $ D_\infty (A) = \sum_{\bi} |\det(A_{\bi})|$. The sum spans over $\binom np$ of
$p$-tuples of indexes $\bi=(1\leq i_1<\cdots< i_p\leq n)$. For any $n\times p$ matrix $A$, the sub-matrix $A_\bi$ is the $p\times p$ matrix containing the rows $\bi= (i_1,\ldots,i_p)$ of $A$. The
second form uses a $L_2$ norm and the corresponding $D$ is $ D_2(A) = (\det A^\top A)^{1/2}$ (the product of singular values).
According to our experiences, these two canonical forms often yield similar results in practical
applications and $D_2$ is less more computational expensive than $D_\infty$.
Interested Readers are referred to
\cite{hannig2015review} for the exact assumptions under which \eqref{eq:overall.fiducial} holds.

Suppose $\Rtheta$ follows the generalized fiducial density $r(\btheta; \by)$.  When using GFI to solve an inference problem, very often one seeks to evaluate the expectation of a function $h(\Rtheta)$, which is defined as
\begin{align}
\label{eq:GFExpectation}
E [h(\Rtheta)\,|\,\by] = \int_{\bTheta} h(\btheta') r(\btheta' ; \by) d\btheta'.
\end{align}
We are guilty of a committing a small abuse of notation in \eqref{eq:GFExpectation}. The expectation is computed with using a generalized fiducial density and not a conditional density. However, just like a conditional expectation it is a measurable function of the observed data. 


As an illustration, consider the following example.  Suppose $p>1$ and it is of interest to compute the marginal generalized fiducial distribution of the first entry of $\btheta$, say $\theta_1$.  One can consider the expectation of the indicator function $h_{t}(\btheta) = 1\{\theta_1 \le t\}$.  The (generalized fiducial) expectation would yield
\begin{align}
\label{eq:fiducial_cdf_expectation}
E [h(\Rtheta)\,|\,\by] = \int_{\theta_1' \le t} r(\btheta' ; \by)
d\btheta' = P({\vartheta}_1 \le t | \by) \overset{\text{def}}{=} R_1(t).
\end{align}
This formulation will be useful to construct interval estimates of $\theta_1$. For example, a lower 95\% confidence interval could be obtained by inverting the function $R_1$ in \eqref{eq:fiducial_cdf_expectation} at $0.95$.  Also, a two-sided 95\% confidence interval could be similarly evaluated by inverting $R_1$ at $0.025$ and $0.975$.  We remark that \eqref{eq:fiducial_cdf_expectation} and more generally \eqref{eq:GFExpectation} cannot be easily computed for most practical problems, and could be much more challenging for massive data sets.

It is very often of interest to provide an measure to summarize the evidence in the data $\by$ supporting the truthfulness of an assertion $A \subset \bTheta$ of the parameter space.  GFI provides a straightforward way to express the amount of belief by the generalized fiducial distribution function:
\begin{align}
  R(A) = E [1\{\Rtheta \in A\} \,|\, \by] = \int_A r(\btheta' ; \by) d\btheta'. \label{eq:FiducialMeasure}
\end{align}
This $R$ function is a valid probability measure and, in many ways, could be viewed as a function similar to posterior distribution in the context of Bayesian inference.


\section{Massive Data Problems}
\label{sec:massive}
For massive data problems where $n$ is huge, the generalized fiducial density in~\eqref{eq:overall.fiducial} could be difficult to evaluate or to obtain samples from.  As mentioned before, one way to address this issue is to partition the whole data set $\bY$ into $K$ subsets $\{\bY_k\}_{k=1}^K$.  For each $k$ the elements of $\bY_k$ are specified by an (nonempty) index set $I_k$ via $\bY_k = \{Y_i, i\in I_k\}$, where $\{I_k\}_{k=1}^K$ form a partition of $\{1,
\ldots, n\}$. 
From~\eqref{eq:overall.fiducial} the generalized fiducial density of $\btheta$ based on observations $\bY_k = \by_k$ for the $k$-th partition is given by
\begin{align} \label{eq:subset.fiducial}
    r_k(\btheta ; \by_k)  = \frac{ f(\by_k ; \btheta) J(\by_k, \btheta)}
    {\int f(\by_k ; \btheta') J(\by_k, \btheta') d\btheta'}
    \defeq \frac{1}{c(\by_k)} f(\by_k ; \btheta) J(\by_k, \btheta).
\end{align}
Let $n_k$ be the size of $I_k$.  It is assumed that for all $k$, $n_k$ is small enough so that samples of $\btheta$ can be generated from \eqref{eq:subset.fiducial} using one single worker.

Combining \eqref{eq:overall.fiducial} and \eqref{eq:subset.fiducial}, the overall generalized
fiducial density $r(\btheta ; \by)$ for the whole observed data set $\by$ can be expressed as a
product of generalized fiducial density $r_k(\btheta ; \by_k)$ and the weights $\prod_{j \ne k}
f(\by_j; \btheta)$:
\[
r(\btheta ; \by)  \propto \frac{J(\by; \btheta)}{J(\by_k; \btheta)} \prod_{j \ne k} f(\by_j;
\btheta)  r_k( \btheta ; \by_k).
\]
This formula decomposes the full density $r(\btheta ; \by)$ into parts of smaller densities $r_k(\btheta ; \by_k)$'s.  With this formula, in below we develop an algorithm to draw samples from the full density $r(\btheta ; \by)$ efficiently by drawing (reweighed) samples from those smaller densities $r_k(\btheta; \by_k)$'s.  Ultimately, these samples will be used to approximate the generalized fiducial measure $R(A)$ defined in~\eqref{eq:FiducialMeasure}.

\ignore{
Suppose $A \subset \bTheta$ be an assertion set of the parameter space. We are going to present a
way to evaluate the fiducial measure $R(A)$ in a tractable manner. The proposing algorithm makes use
of the ``divide and conquer'' philosophy and is parallelizable for big and massive data problems. We
start by dividing the data set into $K$ manageable subsets, $\by_1,\ldots,\by_K$ of size
$n_1,\ldots, n_K$ respectively.  We assume that for each subset $k=1,\ldots, K$ we can generate a
sample $\btheta_{k,t},\ t=1,\ldots, T$ from the GFD \eqref{eq:overall.fiducial} based on the subset
of the data $\by_k$. In next section we derive a computationally feasible formula for combining
these samples into one sample from the GFD based on the whole sample. We will do this using the
importance sampling algorithm \cite{RobertCasella2004}. Figure~\ref{fig:dq} shows a simplified
work flow chart of the algorithm.

\begin{figure}[ht]
\begin{center}
\includegraphics{dq}
\end{center}
\caption{A work flow chart of divide and conquer algorithm}
\label{fig:dq}
\end{figure}
}

\subsection{Importance Sampling}
Importance sampling is a general technique for approximating the expectation of a target distribution via the use of a proposal distribution \citep[e.g., see][]{geweke1989bayesian}.  This subsection develops a naive version of importance sampling to approximate the generalized fiducial measure $R(A)$.  The next subsection will discuss methods for improving this naive version.

For the moment consider using the subset density $r_k(\btheta; \by_k)$ as the proposal.  An advantage of using $r_k(\btheta, \by_k)$ is that, it only requires a subset of of data, and therefore $\by_k$, it would be computationally more feasible than sampling from the original generalized fiducial density $r(\cdot ; \by)$ based on the whole data set $\by$.


Next, for each $k$, define a (un-normalized) proposal density function for $r(\btheta, \by)$ as
\begin{align}
  \pi_k(\btheta) = r_k(\btheta ; \by_k) \label{eq:proposal}.
\end{align}
A normalized version of $\pi_k(\btheta)$ will then be used as the proposal distribution in the importance sampling algorithm.  As similar to most Bayesian problems, MCMC techniques are often employed to obtain samples from this proposal.

Assume now we are able to draw $T$ samples from $\pi_k(\btheta)$ for each $k$.  Denote the samples as $\{\btheta_{k,t}\}$ for $k=1,\ldots,K$ and $t=1\ldots,T$.  Also, for each $k$, define the importance weight function as
\begin{align}
   w_k(\btheta) = \frac{r(\btheta ; \by)}{\pi_k(\btheta)}  =
\frac{J(\by, \btheta)}{J(\by_k,\btheta)}
\prod_{j\ne k} f(\by_j;\btheta).
\label{eq:practical_weight1}
\end{align}
Using those samples $\{\btheta_{k,t},t=1,\ldots,T\}$ generated from the $k$-th subset, one can estimate $R(A)$ by $\hat R_k(A)$ via the usual important sampling method:
\begin{align}
     \hat R_k(A) = \frac{\sum_{t=0}^{T } 1\{\btheta_{k,t} \in A\}
   w_k (\btheta_{k,t})}{\sum_{t=0}^{T} w_k(\btheta_{k,t})}.
  \label{eq:Rhatk}
\end{align}
Combining all the $\hat R_k(A)$'s, one obtains the following improved estimate for $R(A)$:
\begin{align}
   \hat R(A) = \frac{1}{K} \sum_{k=1}^K \hat R_k(A).
  \label{eq:importance.finite}
\end{align}

Consistency of $\hat R_k(A)$ to $R(A)$ can be obtained by an application of the central limit theorem as $T \to \infty$.  This result is presented in Proposition~\ref{prop:1}.  In sequel we assume that $K$ is fixed.

\begin{prop} \label{prop:1}
If the chain $\{\btheta_{k,t}\}$ satisfies Assumption D1 in the appendix and $E \left[w_k(\Rtheta) | \by \right]$ is finite, then the central limit theorem holds for $\hat R_k(A)$; i.e.,
\begin{align*}
  \sqrt{T} [\hat R_k(A) - R(A)] | \by \cid N(0,\sigma^2_k) \quad \text{ as $T \to \infty$},
\end{align*}
where
$ \sigma^2_k = a_k - 2R(A) c_k + R^2(A) b_k$, and
\begin{align*}
  a_k &= Var_{\pi_k}[1\{\btheta_{k,0} \in A\} w_k(\btheta_{k,0})] + 2 \sum_{t=1}^{\infty}
  Cov_{\pi_k} [1\{\btheta_{k,0} \in A\}
  w_k(\btheta_{k,0}), 1\{\btheta_{k,t} \in A\} w_k(\btheta_{k,t})] < \infty,\\
  b_k &= Var_{\pi_k}[w_k(\btheta_{k,0})] + 2 \sum_{t=1}^{\infty} Cov_{\pi_k} [
  w_k(\btheta_{k,0}), w_k(\btheta_{k,t})]< \infty,\\
  c_k &= \sum_{t=0}^\infty Cov_{\pi_k}[ 1\{\btheta_{k,0} \in A\} w_k(\btheta_{k,0}), w_k(\btheta_{k,t})]
  = \sum_{t=0}^\infty Cov_{\pi_k}[1\{\btheta_{k,t} \in A\} w_k(\btheta_{k,t}),w_k(\btheta_{k,0})]<
  \infty.
\end{align*}
\end{prop}
The proof follows the arguments from \cite{geweke1989bayesian} and \cite{jones2004markov}, and is
hence omitted to save space.  This proposition guarantees that $\hat R_k(A)$ is a good approximation
of $R(A)$.  Furthermore, by averaging the $\hat R_k(A)$'s, the variability from the MCMC samples in
$\hat R(A)$ is further reduced, resulting in an more accurate estimate for $R(A)$.

\subsection{Improving Importance Weights}
Amongst other factors, the overall speed of the above importance sampling algorithm relies on how
fast one could compute the weights~\eqref{eq:practical_weight1}.  The first term $J(\by,
\btheta)/J(\by_k,\btheta)$ is the lengthy term to compute, as it involves the whole data set $\by$.
Notice that by law of large numbers this term behaves almost like a constant w.r.t.~$\btheta$ when comparing to the second
and dominating term $\prod_{j\ne k} f(\by_j; \btheta)$; this is particularly true when $\by_k$ is
a representative sample of $\by$.  Motivated by this, we propose approximating the original weight
function~\eqref{eq:practical_weight1} by ignoring the first term, which gives the following improved
weight function
\begin{align}
\label{eq:practical_weight2}
\tilde{w}_k(\btheta) =  \frac{J(\by_k, \btheta)}{J(\by,\btheta)}
w_k(\btheta) =\prod_{j\ne k} f(\by_j; \btheta).
\end{align}
With this $R(A)$ can be estimated, in a similar fashion as in~\eqref{eq:Rhatk}, with
\begin{align}
\label{eq:tilde_Rk}
     \tilde R_k(A) = \frac{\sum_{t=0}^{T } 1\{\btheta_{k,t} \in A\}
   \tilde{w}_k (\btheta_{k,t})}{\sum_{t=0}^{T} \tilde{w}_k(\btheta_{k,t})}.
\end{align}
We have the following proposition immediately.

\begin{prop}
\label{prop:2}
If the chain $\{\btheta_{k,t}\}$ satisfies Assumption~D1 in the appendix and if
  $E\left[ \frac{\tilde{w}_k^2(\Rtheta)}{ w_k(\Rtheta) } | \by \right]$ is
  finite,
then
\begin{align*}
  \sqrt{T} [\tilde R_k(A) - R_k^*(A)] | \by \cid N(0,\sigma^2_k) \quad \text{ as $T \to \infty$},
\end{align*}
where
\begin{align*}
R_k^*(A) & = E\left[1\{\btheta_{k,t} \in A\} \frac{J(\by_k, \Rtheta)}{J(\by, \Rtheta)} | \by
\right]
\Big/ E\left[ \frac{J(\by_k, \Rtheta)}{J(\by, \Rtheta)} | \by \right], \\
\sigma^2_k & = (a_k - 2R_k^*(A) c_k + R_k^*(A)^{2} b_k) \Big/
{\left\{ E\left[ \frac{J(\by_k, \Rtheta)}{J(\by, \Rtheta)} | \by \right]\right\}}^2,
\end{align*}
and
  $a_k$, $b_k$ and $c_k$ are defined in Proposition~\ref{prop:1}.
\end{prop}
The major idea behind the proof of Proposition~\ref{prop:2} is very similar to that of Proposition~\ref{prop:1}, and therefore is omitted for brevity.  This proposition indicates that $\tilde R_k(A)$ is converging to $R_k^*(A)$ as $T \to \infty$ and hence $\tilde R_k(A)$ is a biased estimator of $R(A)$.  This bias is introduced when $w_k(\btheta)$ are replaced by $\tilde{w}_k(\btheta)$ in order to obtain higher computational speed.

The next proposition shows that the bias in $\tilde R_k(A)$ is asymptotically negligible, providing a theoretical support of the use of $\tilde{w}_k(\btheta)$.  The convergence in probability below is with respect to the distribution of the data $\by$. The proof is given in the appendix.

\begin{prop}
\label{prop:3}
Let $\hat \btheta_n$ be the maximum likelihood estimate of $\btheta$.  Suppose Assumptions E1 and E2
in the appendix hold.  Then as $n \to \infty$,
\begin{align}
\label{eq:prop3.2}
  \sqrt{n}E\left[  \left| \frac{J(\by_k, \Rtheta)}{J(\by, \Rtheta)} -
  \frac{J(\by_k, \hat \btheta_n)}{J(\by, \hat \btheta_n)} \right| \Big| \by \right] \cip 0
\end{align}
and
\begin{align}
  R_k^*(A) = R(A) + o_p(n^{-1/2}) . \label{eq:prop3.3}
\end{align}
\end{prop}

Now we are ready to present our main theoretical result.
\begin{prop}
\label{prop:4}
Under the conditions of Propositions~\ref{prop:2} and~\ref{prop:3},  we have, for all $\varepsilon>0$,
\begin{align*}
  P\left\{\sqrt{n} | \tilde R_k(A)-  R(A) | > \varepsilon \,\Big|\, \by \right\} \cip 0
\end{align*}
 as $T \to \infty$, $n \to \infty$ and $n/T \to 0$.
\end{prop}
Proposition \ref{prop:4} indicates that the fiducial probability of an assertion set $A \subset
\bTheta$ can be approximated by $\tilde R_k(A)$ with high accuracy.  Note that this asymptotic
result holds when both $n$ and $T$ go to infinity, with $T$ goes much faster than $n$ does.  These
conditions are required since we have to ensure that the approximation error due to the importance
sampling procedure is comparable to the bias introduced in the weight function
$\tilde{w}_k(\btheta)$ in~\eqref{eq:practical_weight2}.  The proof of this proposition is given in
the appendix.

Since from Proposition~\ref{prop:4} each $\tilde R_k(A)$ is a consistent estimator, it is natural to define our final estimator of $R(A)$ as $\tilde R(A)$:
\begin{align}
   \tilde R(A) = \frac{1}{K} \sum_{k=1}^K \tilde R_k(A).
  \label{eq:importance.finite2}
\end{align}
Note that the averaging operation further reduces the variability in the estimation in $\tilde R(A)$.  Once $\tilde R(A)$ is obtained, it can be used to conduct statistical inference about the parameters of interest, in a similar manner as with a posterior distribution in the Bayesian context.

\subsection{Practical Algorithms}
This subsection presents two practical algorithms that implement the above results.  The first one is a straightforward and direct implementation, and is listed in Algorithm~\ref{alg:direct}.
\begin{algorithm}
\caption{Direct Implementation}
\label{alg:direct}
\begin{enumerate}
\item Partition the data $\by$ into $K$ subsets $\by_1, \ldots, \by_K$.  Each subset $\by_k$ becomes the input for one of the $K$ parallel jobs.
\item For $k=1, \ldots, K$, the $k$-th worker generates a sample of $\Rtheta$ of size $T$ from
\eqref{eq:proposal} and returns the result to the main node.
\item The collected samples are broadcasted to all workers and each worker computes its relevant
portion of $\tilde{w}_k(\Rtheta)$ in \eqref{eq:practical_weight2} and returns the result to the
main node.

\item Combine the results from the workers to obtain $\tilde{w}_k(\Rtheta)$ and calculate $\tilde R_k(A)$ using \eqref{eq:tilde_Rk}.

\item Average all the $\tilde R_k(A)$'s and obtain the final estimate $\tilde R(A)$ as in \eqref{eq:importance.finite2}.
\end{enumerate}
\end{algorithm}

The effectiveness of Algorithm~\ref{alg:direct} depends on the importance weights and the effective
sample sizes of the the importance samplers. For an implementation of $K$ workers and each worker
stores $n_k$ observations, the relative efficiency \citep{kong1992note,liu1996metropolized}
for each worker is approximately
\begin{align*}
\frac{1}{E_{\pi_k}\tilde{w}^2_k(\Rtheta)} \approx O\{\exp(- \tau K)\},
\end{align*}
where the constant $\tau$ depends on the likelihood being considered. For large values of $K$, the
corresponding number of fiducial samples has to increase exponentially in order to achieve the
same estimation accuracy. To address this issue, we propose a modified algorithm, which is given as Algorithm~\ref{alg:improved}.


\begin{algorithm}
\caption{Improved Implementation: \aname}
\label{alg:improved}
\begin{enumerate}
\item Partition the data $\by$ into $K$ subsets $\by_1, \ldots, \by_K$.  Each subset $\by_k$ becomes the input for one of the $K$ parallel jobs. Here $K$ is chosen as a power of 2.
\item For $k=1, \ldots, K$, the $k$-th worker generates a sample of $\Rtheta$ of size $T$ from
\eqref{eq:proposal} and returns the result to the main node.
\item Repeat the following until one subset is left:
	\begin{enumerate}
    	\item For any two subsets, say $k_i$ and $k_j$,
        	\begin{enumerate}
            	\item Compute parallelly the weights as
                	$\tilde{w}_{k_i}(\Rtheta) = f(\by_{k_j};\Rtheta).$
                 \item Return the weights to the main node.

                 \item At the main node, resample the sample of $\Rtheta$ from subset $k_i$ with weights $\tilde{w}_{k_j}$ and resample the sample of $\Rtheta$ from subset $k_j$ with weights $\tilde{w}_{k_j}$.
                 \item Merge the two samples in the previous step into form a single sample of $\Rtheta$.
                   \item Group the subsets $\by_{k_i}$ and $\by_{k_j}$ together. 
             \end{enumerate}
         \item Repeat (a) with another pair of subsets until there are only half the original number of subsets remain.
    \end{enumerate}
    \item With the combined sample of $\Rtheta$, compute $\tilde R(A)$ by using
    $\tilde R_k(A) = T^{-1} \sum_{t=0}^{T } 1\{\Rtheta_{k,t} \in A\}$ and \eqref{eq:importance.finite2}.
\end{enumerate}
\end{algorithm}

With Algorithm~\ref{alg:improved}, the effective number of workers is reduced to $\log K$ and the relative efficiency of the importance samplers increases from $O\{\exp(- \tau K)\}$ to $O(K^ {-1})$. Algorithm~\ref{alg:improved} also reduces the number of evaluations of the likelihood functions since the evaluations are now only required by the merging fiducial samples, whereas in Algorithm~\ref{alg:direct}, the
evaluations are required by each of the the fiducial samples. 

In all the numerical work to be reported below, only Algorithm~\ref{alg:improved} was used.



\section{Simulations}
\label{sec:simulation}
To investigate the feasibility and the empirical performance of the proposed approach, we consider simulated
data from two different models: a mixture of two normal distributions and a linear regression
model with $p$ covariates and Cauchy distributed errors. For each of the models, we
will construct fiducial confidence intervals for the parameters. Their nominal and
empirical converges will be presented. We will vary the simulation settings using different sample sizes
$n$, different number of workers $K$, and also different number of parameters.


In each simulation setup, we first generate $n$ observations from the underlying model and then
divide them randomly into $K$ groups. Each group of observations will be sent to a parallel worker
for further processing. Each of the $K$ workers will then perform a MCMC procedure to sample from $T$
particles using~\eqref{eq:proposal}. In our simulation, the Metropolis Hastings algorithm is
implemented for this purpose and $T$ is chosen to be 10,000 for all cases. Each setting is then
repeated 100 times to obtain the empirical converges for the one sided fiducial confidence
intervals.


\subsection{Mixture of Normals}
  The density of $Y$ is $f_Y(y) = \gamma \phi(y; \mu_1, \sigma) + (1-\gamma) \phi(y; \mu_2, \sigma)$,
  where $\phi (y;\mu,\sigma)$ is the normal density with mean $\mu$ and variance $\sigma^2$. For
  simplicity, we assume that $\sigma=1$ is known. The true values of $(\mu_1, \mu_2, \gamma)$ is $
  (-1,1,0.6)$. Note $\mu_1 < \mu_2$ so identifiability is ensured. Three values of $n=10^5$, $2\times 10^5$ and $4\times 10^5$, and six values of $K=1, 2, 4, 6, 16$ and $32$ are used.

\begin{figure}[ht]
\begin{center}
\includegraphics[scale=0.9]{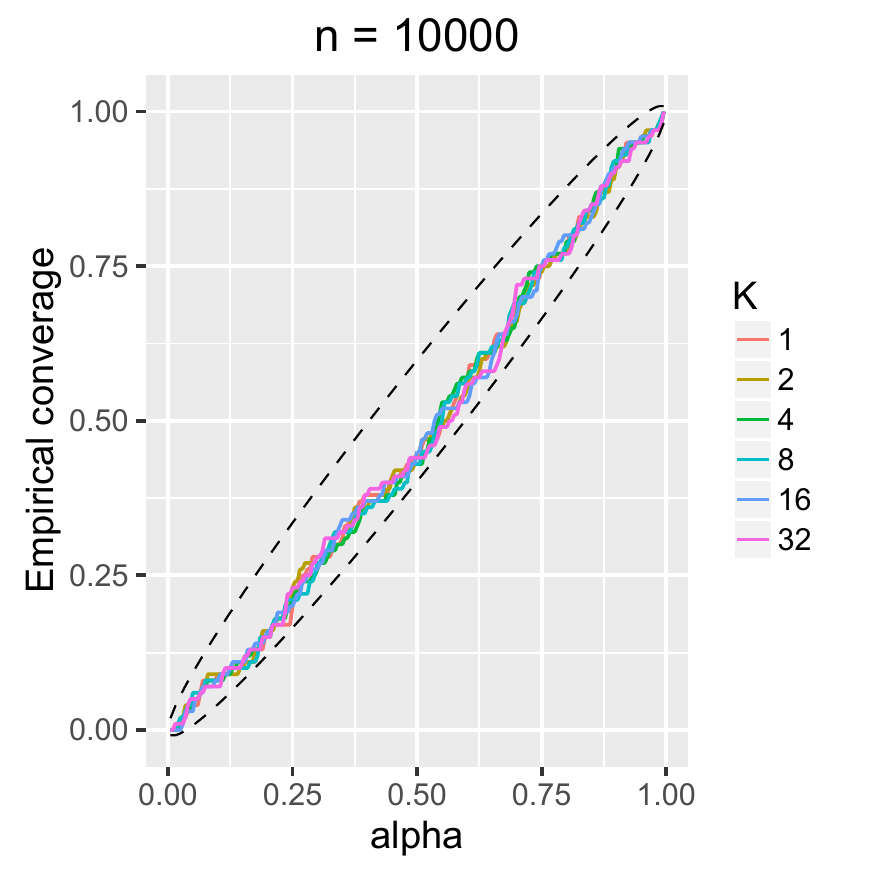}
  \includegraphics[scale=0.9]{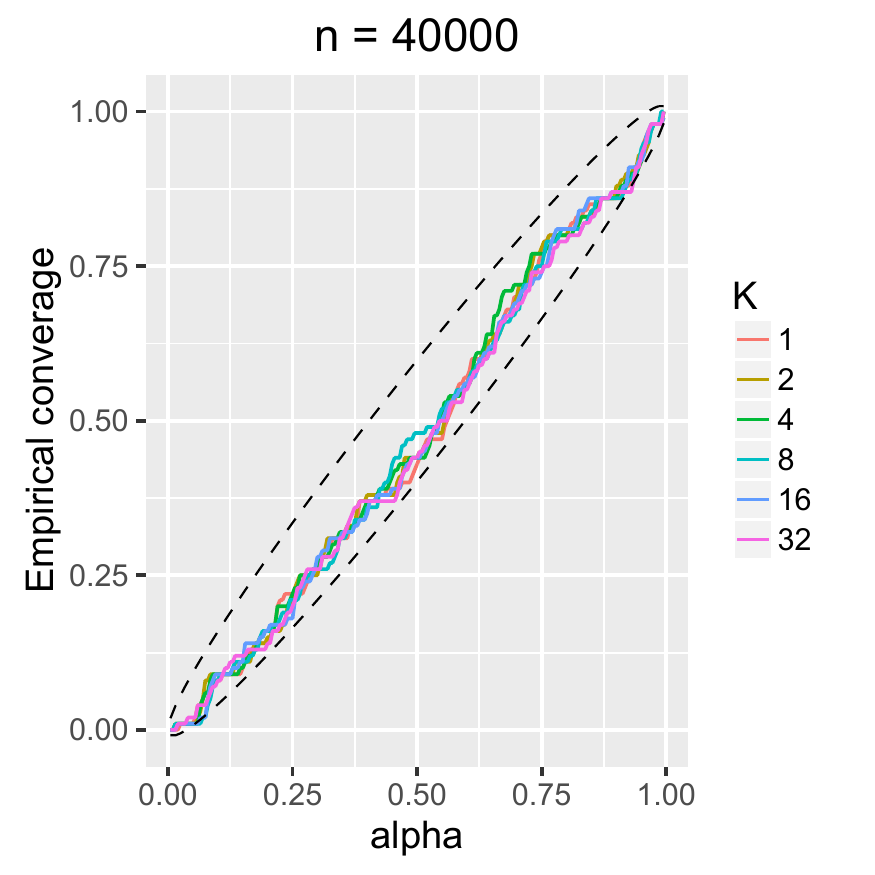}
  \end{center}
  \caption{The empirical converges for the lower sided fiducial confidence intervals for $\mu_1$
  for different number of observations and number of workers. The results for $n=2\times 10^5$ and $\mu_2$ are similar and hence omitted.}
  \label{fig:mixture1}
  \end{figure}
\begin{figure}[ht]
\begin{center}
\includegraphics[scale=0.9]{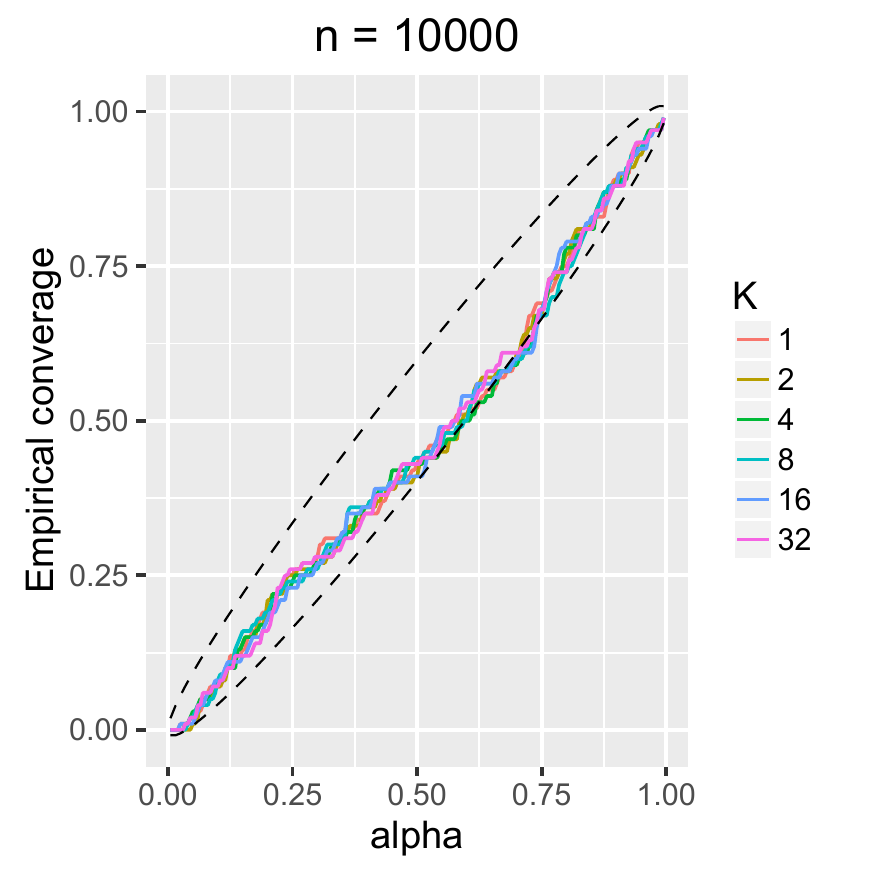}
  \includegraphics[scale=0.9]{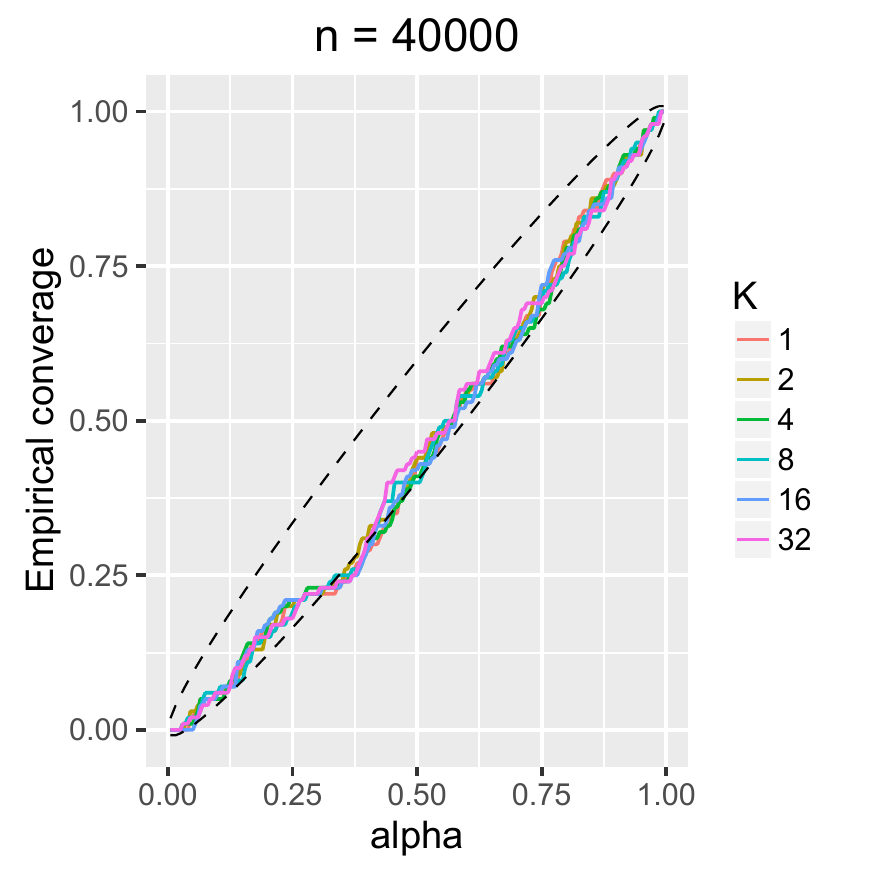}
  \end{center}
  \caption{Similar to Figure~\protect\ref{fig:mixture1} but for the parameter $\gamma$.}
  \label{fig:mixture2}
  \end{figure}

For the cases $n=10^5$ and $n=4\times 10^5$, the empirical converges for all $100(1-\alpha)\%$ lower sided fiducial confidence intervals for the parameters $\mu_1$ and $\gamma$ are shown, respectively, in Figure~\ref{fig:mixture1} and Figure~\ref{fig:mixture2}. The dotted lines are the theoretical confidence interval for the empirical coverages: $\alpha \pm 1.96 \sqrt{\alpha (1-\alpha)/100}$.  From these figures one can see that the proposed method performs very well with empirical coverages agreeing with the nominal coverages at all levels.

\subsection{Cauchy Regression}
  The model is $Y = \beta_0 + \bbeta \bX + \sigma W$, where $\beta_0 \in \Re$, $\bbeta \in \Re^p$
  and $\sigma >0$. The error distribution of $W$ is standard Cauchy and the design matrix $\bX$ is multivariate normal with zero mean, unit variance and pairwise correlation $\rho=0.1$. The following parameter values are used:
  $\sigma=1$, $\beta_0=0$ and $\bbeta=(\beta_1,\beta_2,\beta_3,\beta_4,\beta_5,\ldots)=(1,1,1,0,0,\ldots)$; i.e., all slope coefficients are zero except the first three. The number of observations is fixed at $n=10^5$ while $K=1,2,4,8$ and $16$, and $p=5,7$ and $10$.

\begin{figure}[ht]
\begin{center}
\includegraphics[scale=0.9]{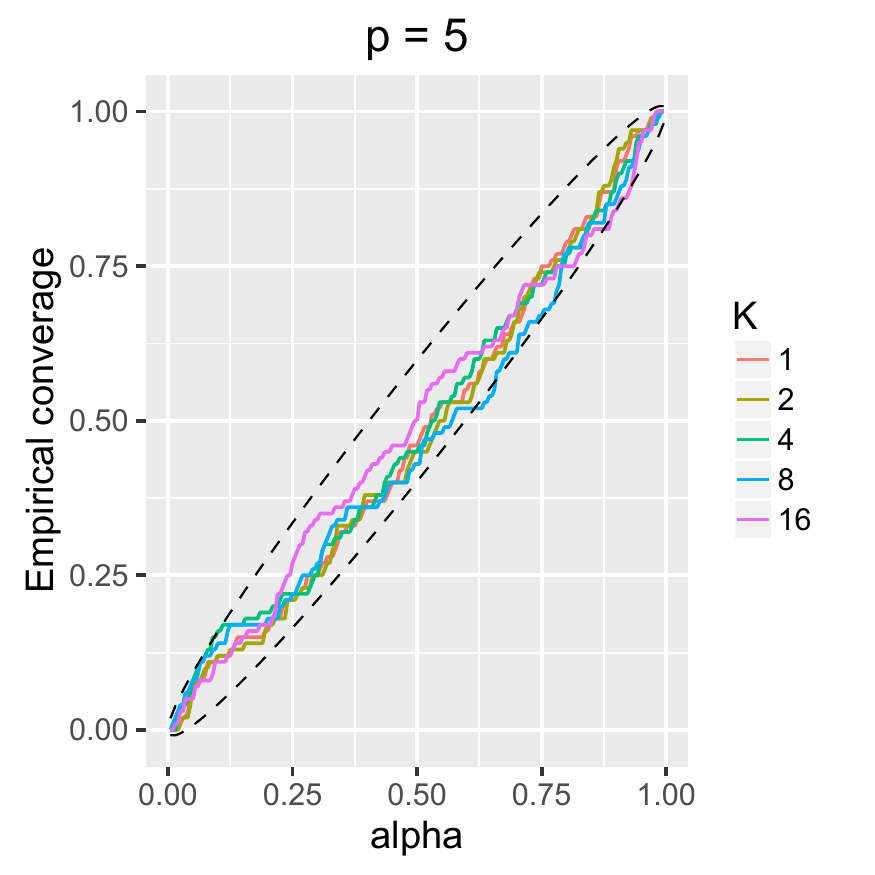}
  \includegraphics[scale=0.9]{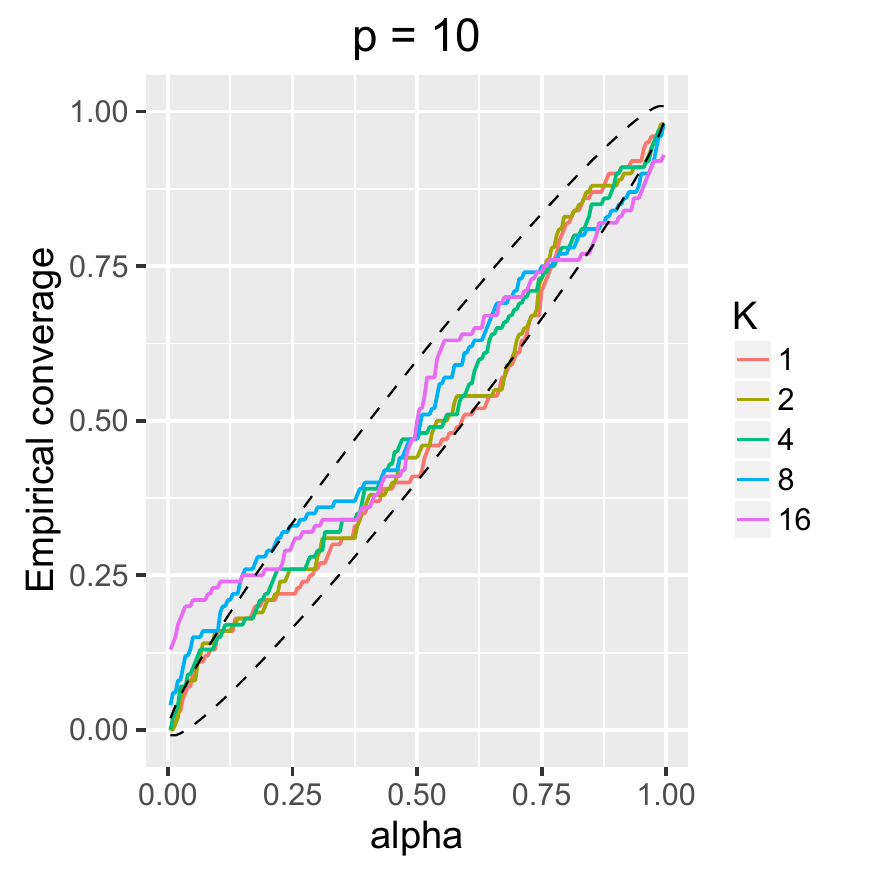}
\end{center}
\caption{The empirical converges for the lower sided fiducial confidence intervals for $\beta_1$ for different number of covariates and number of workers. The results for $p=7$ and for $\beta_2$ and $\beta_3$ are similar and hence omitted.}
  \label{fig:cauchy1}
  \end{figure}

\begin{figure}[ht]
\begin{center}
\includegraphics[scale=0.9]{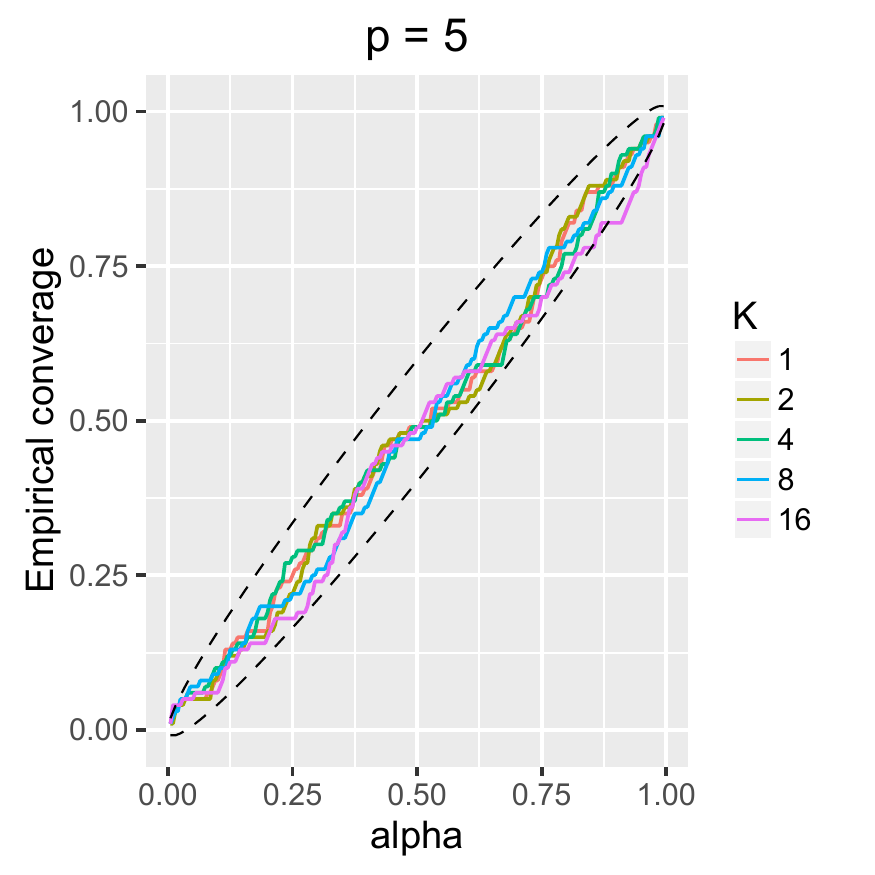}
  \includegraphics[scale=0.9]{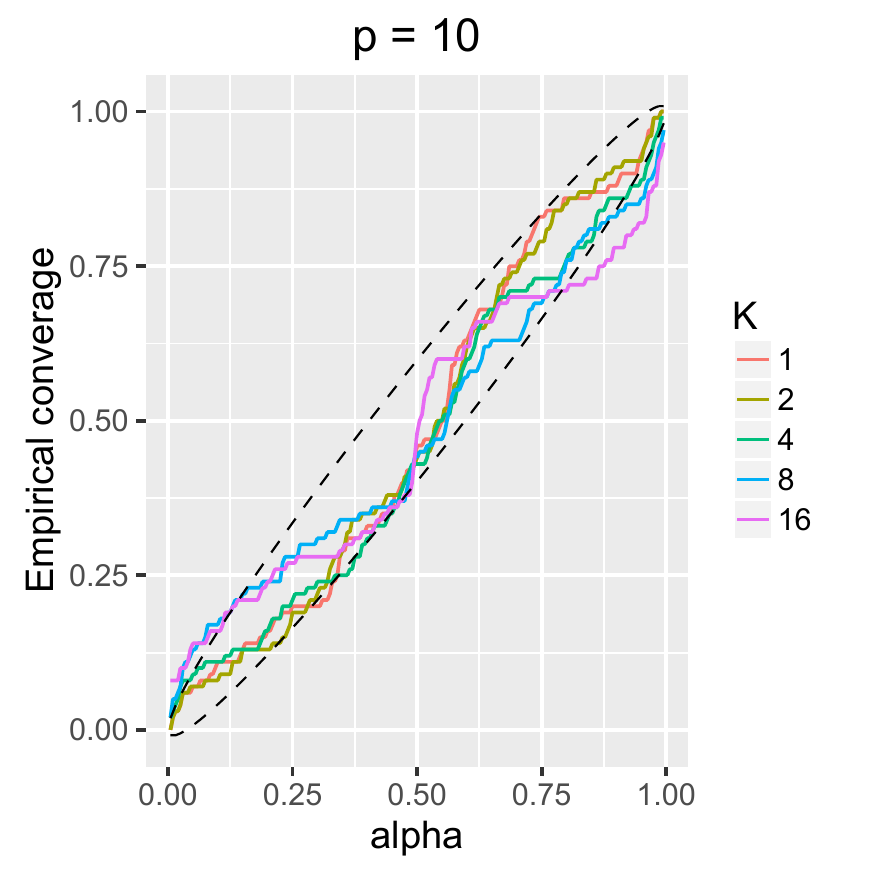}
\end{center}
\caption{Similar to Figure~\protect\ref{fig:cauchy1} but for the parameter $\beta_4$. The results for $\beta_j$, $j>4$ are similar and hence omitted.}
  \label{fig:cauchy4}
  \end{figure}

For the cases of $p=5$ and $p=10$, the empirical converges for the $100(1-\alpha)\%$ lower sided fiducial confidence intervals for the parameters $\beta_1$ and $\beta_4$ are shown, respectively, in Figure~\ref{fig:cauchy1} and Figure~\ref{fig:cauchy4}. Similar to the previous subsection, the dotted lines are the theoretical confidence interval for the empirical coverages: $\alpha \pm 1.96 \sqrt{\alpha (1-\alpha)/100}$.  As with the previous subsection, the proposed method produced very good results.

\subsection{Computational Efficiency}

\begin{figure}[ht]
\center
\includegraphics[scale=0.9]{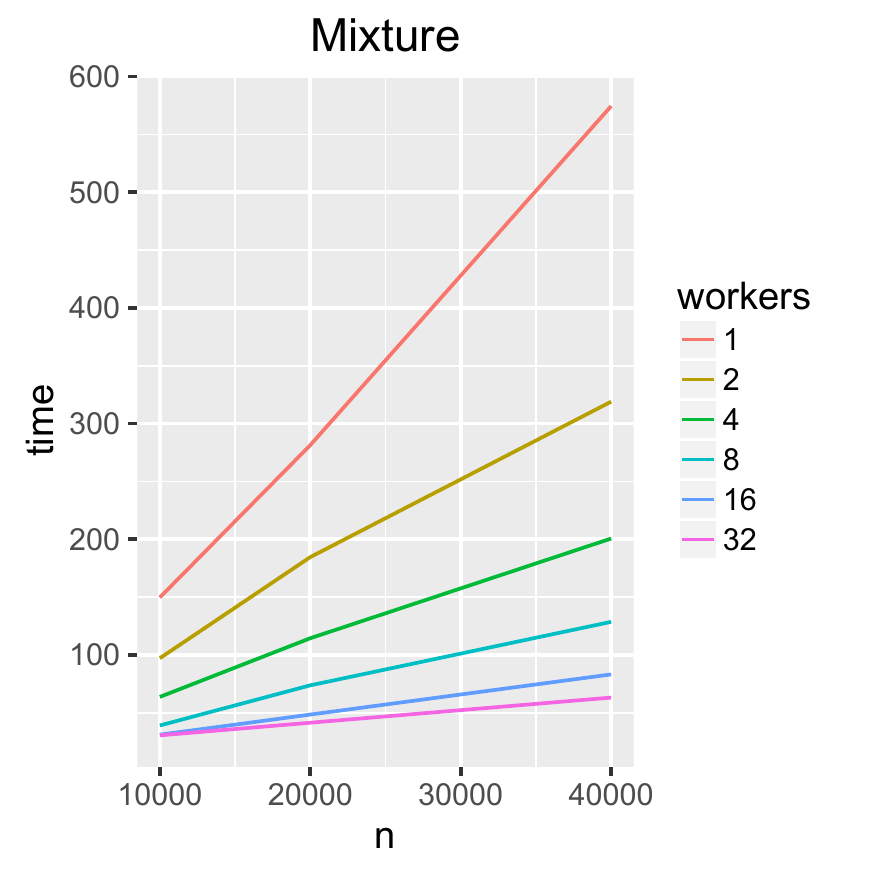}
\includegraphics[scale=0.9]{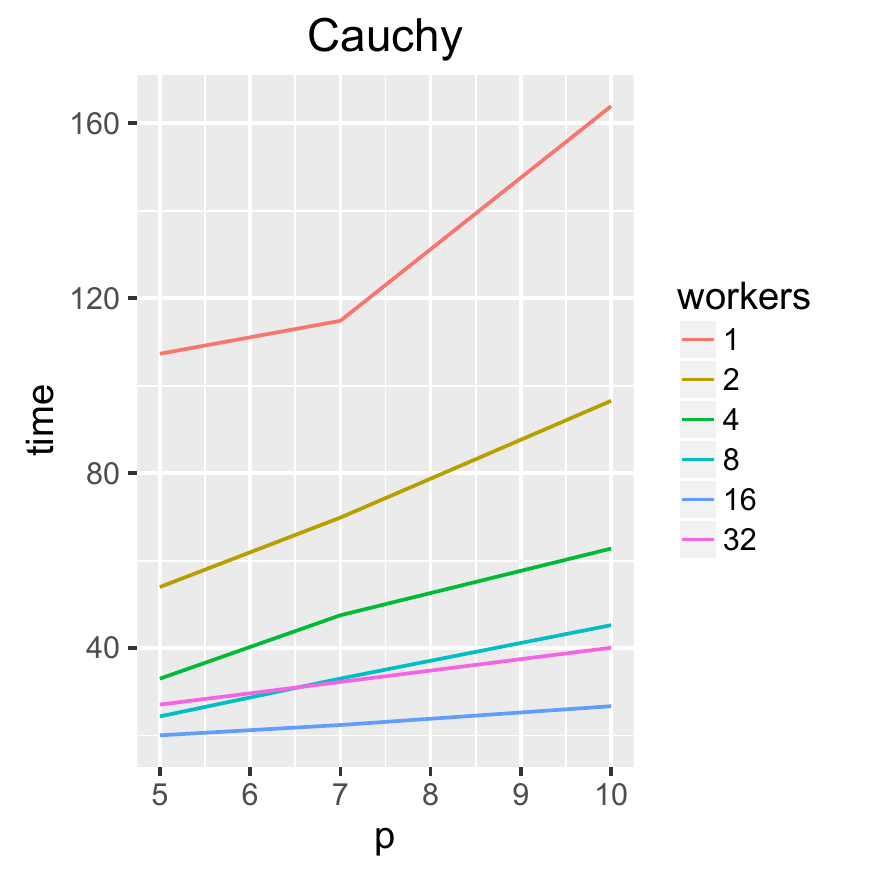}
  \caption{The elapsed time required for the normal mixture and Cauchy models for different number of
  workers $K$.}
  \label{fig:time}
  \end{figure}

One primary goal of this article is to develop a scalable solution to reduce the computational time required in performing generalized fiducial inference. The computational times in the above normal mixture and Cauchy models
are reported in Figure~\ref{fig:time}. It can be seen that the proposed method is more time-efficient when the sample size (for the normal mixture model) or the number of covariates (for the Cauchy model) increases.  

Intuitively, one would believe that more workers would lead to a larger reduction of computational time.  This is partially true, as if the number of workers exceeds the maximum beneficial optimal value, the total computational time and statistical optimality may rebound; see \cite{cheng2015computational} for an interesting discussion. A major reason is that there is a trade-off between the actual computation cost and the cost for data transfer and communication among the workers in this divide-and-conquer strategy. For the Cauchy example, the total computational cost for the case of $32$ workers was ``unsurprisingly'' more than that of the case of $16$ workers. It is probably because more time was spent in data manipulation and allocation than in the real computations.

In the above simulation studies, different models and different error distributions were carefully chosen with the hope to represent most of the practical scenarios. However, just as any other simulation studies, the above numerical experiments by no means are sufficient to cover all cases that one may encounter in practice, and therefore caution must be exercised in drawing conclusions from the empirical results. Despite this, the following conclusion is appropriate. GFI can be made parallelized to handle massive data problems with Algorithm~\ref{alg:improved} and the resulting statistical inferences are asymptotically correct. The performance of Algorithm~\ref{alg:improved} depends on the total sample size and worker sample sizes. Simulation results show that the fiducial confidence intervals produced by the algorithm have very reliable and attractive frequentist properties.

\section{Data analysis: Solar flares}
\label{sec:solar}
In this section the methodology developed above is applied to help understand the occurrences of solar flares.  The data were collected by the instrument Atmospheric Imaging Assembly (AIA) mounted on the Solar Dynamics Observatory (SDO).  As stated in the official NASA website, SDO was designed to help study the influence of the Sun on Earth and Near-Earth space.  SDO was launched on 2010.

\begin{figure}[ht]
\label{fig:sun}
\begin{center}
\includegraphics[scale=0.9]{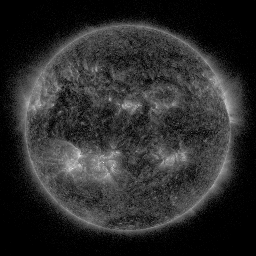}
\includegraphics[scale=0.9]{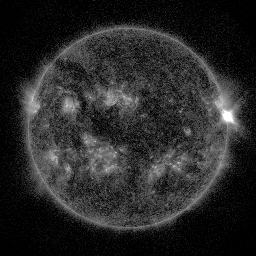}
\end{center}
\caption{Two images of the Sun captured by AIA. There is a solar flare occurring in the right image, near the right end of the equator of the Sun. The white spot intensity has a value of 253 on a 8-bit scale from 0 (black) to 255 (white).}
\end{figure}

The instrument AIA captures images of the Sun in eight different wavelengths every 12 seconds; see Figure~\ref{fig:sun} for two examples.  The image size is $4096 \times 4096$ pixels, which provides a total of 1.5 terabytes compressed data per day.  An uncompressed and pre-processed version of the data can be obtained from \cite{schuh2013large}.  Here each image was partitioned into $64 \times 64$ squared and equi-sized sub-images, each consists of $64 \times 64$ pixels.  For each sub-image, 10 summary statistics were computed, such as the average and the standard deviation of the pixel values.

\begin{figure}[ht]
\includegraphics[scale=0.8]{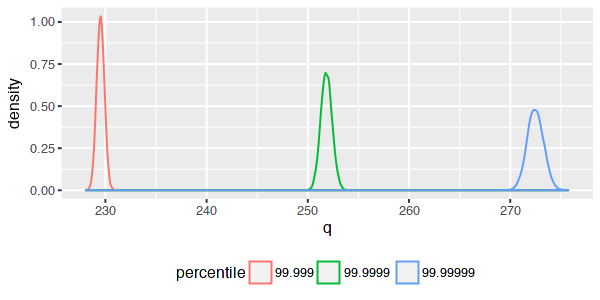}
\caption{Kernel density estimates of the fiducial densities of the brightness of 99.999, 99.9999 and 99.99999 percentiles solar flare events. One can see that an observed value of 253 roughly corresponds to 99.9999\% (1 in a million). Values over 255 indicate events with brightness that is higher than the resolution of the instrument.}
\label{fig:extremedensity}
\end{figure}

\begin{figure}[ht]
\begin{center}
\includegraphics[scale=0.4]{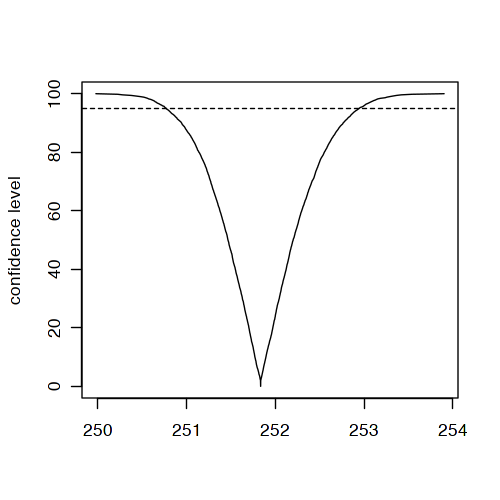}
\includegraphics[scale=0.35]{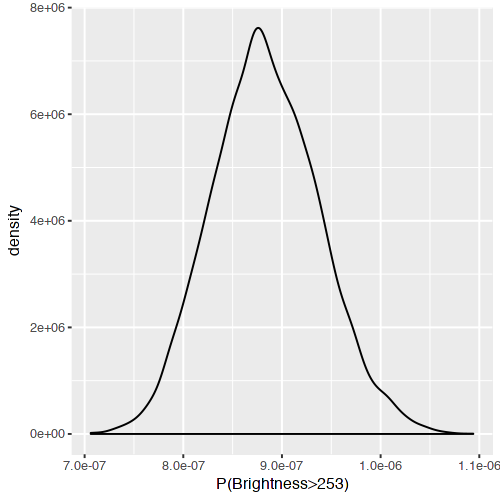}
\end{center}
\caption{Left: Confidence curve for the 99.9999 percentile. Right: Fiducial probability of brightness greater than 253.}
\label{fig:confidence_curve}
\end{figure}

A solar flare is a sudden eruption of high-energy radiation from the Sun's surface, which could cause disturbances on communication and power systems on Earth.  In those images captured by AIA, such solar flares are characterized by extremely bright spots; see the right panel of Figure~\ref{fig:sun} for an example.  \cite{WandlerHannig2012b} provide a solution for estimating extremes using GFI for small data sets. For this large data set we use the averaged pixel values computed from \cite{schuh2013large} and the proposed method to parallelize the computations. Figure~\ref{fig:extremedensity} reports the kernel based estimates for the fiducial densities of the 99.999, 99.9999 and 99.99999 percentiles of the brightness.  These densities can help astronomers determine the frequency, predict the occurrences of the solar flares, and understand the limitations of their instruments. Figure~\ref{fig:confidence_curve} displays the confidence curve for the 99.9999 percentile for the solar flare brightness. The 95\% confidence interval is (250.8, 253.0) and a solar flare of brightness in this range is likely to happen with 1 in a million chance. The fiducial probability of brightness greater than 253 is also computed and displayed in Figure~\ref{fig:confidence_curve}.

The simulations were run on UCDavis Department of Statistics computer cluster, each node of the cluster is equipped with a 32-core AMD Opteron(TM) Processor 6272. The program took about 15s to finish the fiducial sample generation process when 32 workers are in work and it took about 80s when only 4 workers are in place.



\section{Conclusion and discussion}
\label{sec:conclude}
In this paper generalized fiducial inference is paired with importance sampling to develop a method for the distributed analysis of massive data sets.  In addition to point estimates, the resulting method is also capable of producing uncertainty measures for such quantities.  Another attractive feature of the method is that it only requires minimal communications amongst workers.  Via mathematical calculations and numerical experiments, the method is shown to enjoy excellent theoretical and empirical properties.

The proposed method assumes the sub-sample in each worker is a random sample from the original data set.  Therefore a useful extension of the current work is to relax this assumption.  Another important extension is to allow for heterogeneity that is a common feature of massive data sets that are obtained from potentially disparate sources. One possible computationally efficient approach for handling this issue was proposed in the ``small data'' inter-laboratory comparison context by \cite{hannig2017fusion}. Their idea seems especially promising in the massive data context if one could ensure that the within each subset is relatively homogeneous while the data between subsets is potentially heterogeneous. 

\section*{Acknowledgement}
The work of Lee was partially supported by National Science Foundation under grants DMS-1512945 and DMS-1513484. Hannig's research was supported in part by the National Science Foundation under Grant No. 1512945 and 1633074.

\appendix
\section{Technical Details}

This appendix provides technical details.

\subsection*{Assumptions}
We begin with a set of assumptions which allow us to work on the theories.

We start by listing the standard assumptions sufficient to prove that the maximum likelihood estimators are asymptotically normal  \citep{LehmannCasella1998}.

\begin{enumerate}
\item [{(A0)}] The distributions $P_{{\btheta}}$ are distinct.
\item [{(A1)}] The set $\left\{ y:f(y|{\btheta})>0\right\} $ is
independent of the choice of ${\btheta}$.
\item [{(A2)}] The data $\bY=\{Y_{1},\dots,Y_{n}\}$ are iid
with probability density $f(\cdot|{\btheta})$.
\item [{(A3)}] There exists an open neighborhood about the true parameter
value ${\btheta}_{0}$ such that all third partial derivatives
$\left(\partial^{3}/\partial\theta_{i}\partial\theta_{j}\partial\theta_k\right)f(\by|{\btheta})$
exist in the neighborhood, denoted by $B({\btheta}_{0},\delta)$.
\item [{(A4)}] The first and second derivatives of $L({\btheta},y)=\log f(y|{\btheta})$
satisfy
\[
E_{{\btheta}}\left[\frac{\partial}{\partial\theta_{j}}L({\btheta},y)\right]=0
\]
and
\begin{eqnarray*}
I_{j,k}({\btheta}) & = & E_{{\btheta}}\left[\frac{\partial}{\partial\theta_{j}}L({\btheta},y)\cdot\frac{\partial}{\partial\theta_k}L({\btheta},y)\right]\\
 & = & -E_{{\btheta}}\left[\frac{\partial^{2}}{\partial\theta_{j}\partial\theta_k}L({\btheta},y)\right].
\end{eqnarray*}

\item [{(A5)}] The information matrix $I({\btheta})$ is positive
definite for all ${\btheta}\in B({\btheta}_{0},\delta)$
\item [{(A6)}] There exists functions $M_{jkl}(\by)$ such that
\[
\sup_{{\btheta}\in B({\btheta}_{0},\delta)}\left|\frac{\partial^{3}}{\partial\theta_{j}\partial\theta_k\partial\theta_{l}}L({\btheta},y)\right|\le M_{j,k,l}(y)\;\;\;\;\textrm{and}\;\;\; E_{{\btheta}_{0}}M_{j,k,l}(Y)<\infty
\]
\end{enumerate}

Next we state conditions sufficient for the Bayesian posterior distribution to be close to that of the MLE  \citep{vanderVaart1998,GhoshRamamoorthi2003}. The prior used is the limiting fiducial prior
Let $\pi({\btheta})=E_{{\btheta}_{0}}J_{0}(Y_{0},{\btheta})$
and $L_{n}({\btheta})=\sum L({\btheta},Y_{i})$
\begin{enumerate}
\item [{(B1)}] For any $\delta>0$ there exists $\epsilon>0$ such that
\[
P_{{\btheta}_{0}}\left\{ \sup_{{\btheta}\notin B({\btheta}_{0},\delta)}\frac{1}{n}\left(L_{n}({\btheta})-L_{n}({\btheta}_{0})\right)\le-\epsilon\right\} \to1
\]

\item [{(B2)}] $\pi\left({\btheta}\right)$ is positive at ${\btheta}_{0}$
\end{enumerate}

Finally we state assumptions on the Jacobian function.  Recall
$\pi({\btheta})=E_{{\btheta}_{0}}J_{0}(X_{0},{\btheta})$.

\begin{enumerate}
\item[(C1)] For any $\delta>0$
\[
\inf_{{\btheta}\notin B({\btheta}_{0},\delta)}\frac{\min_{i=1\dots n}L({\btheta},Y_{i})}{\left|L_{n}({\btheta})-L_{n}({\btheta}_{0})\right|}\stackrel{P_{{\btheta}_{0}}}{\longrightarrow}0
\]
 where $L_{n}\left({\btheta}\right)=\sum_{i=1}^{n}\log f\left(y_{i} ;{\btheta}\right)$
and $B\left({\btheta}_{0},\delta\right)$ is a neighborhood
of diameter $\delta$ centered at ${\btheta}_{0}$.

\item[(C2)]
The Jacobian function $\binom{n}{p}^{-1}J\left(\bY,{\btheta}\right)\stackrel{a.s.}{\to}\pi\left({\btheta}\right)$
uniformly on compacts in ${\btheta}$.
\end{enumerate}

\begin{enumerate}
  \item[(D1)] The MCMC chain $\{\btheta_{k,t}\}$ is an ergodic Markov chain with marginal density
  $\pi_k$
  defined in
  \eqref{eq:proposal} and satisfying at least one of the followings:
  \begin{enumerate}
    \item[(a)]  geometrically ergodic and detailed balanced, or
    \item[(b)] uniformly ergodic.
  \end{enumerate}
  Moreover, if $k\ne k'$, chains from different workers, say $\{\btheta_{k,t}\}$ and $ \{\btheta^{
  (t)}_{k'}\}$, are
  independent given $\by$.

\end{enumerate}

\begin{enumerate}
  \item[(E1)] Let $u_k(\by, \btheta)=\frac{\partial}{\partial \btheta}\frac{J(\by_k, \btheta)}{J(\by,
  \btheta)}$,
   there exists $U(\by)$ s.t. $u_k(\by, \btheta) \le U(\by)$ for all $\btheta$ with probability tending to 1.
  \item[(E2)] $\int_{\Real^p} |t| f_{\sqrt{n}(\Rtheta-\hat \btheta)}(t)
  dt \cip \int_{\Real^p} |t| \phi(t; 0, I^{-1}(\btheta_0)) dt$, where
  $f_{\sqrt{n}(\Rtheta-\hat \btheta)}$ is the scaled generalized fiducial density and $\phi$ is the
  multivariate normal density function.
\end{enumerate}

\subsection*{Proofs}

\begin{proof}[Proof of Proposition 3.]
We first consider
\begin{align*}
  &\sqrt{n}E\left[  \left|  \frac{J(\by_k, \Rtheta)}{J(\by, \Rtheta)} -
  \frac{J(\by_k, \hat \btheta_n)}{J(\by, \hat \btheta_n)} \right|
  \Big| \by \right]\\
  &= \int_\Xi  \sqrt{n} \left|\frac{J(\by_k, \btheta)}{J(\by, \btheta)}-
  \frac{J(\by_k, \hat \btheta_n)}{J(\by, \hat \btheta_n)} \right| r(\btheta) d\btheta \\
  &= \int_{\Real^p}
  \sqrt{n} \left| \frac{J(\by_k, \hat \btheta_n + \frac{t}{\sqrt{n}})}{J(\by, \hat \btheta_n
  + \frac{t}{\sqrt{n}})} - \frac{J(\by_k, \hat \btheta_n )}{J(\by, \hat \btheta_n )}
  \right| f_{\sqrt{n}(\Rtheta-\hat \btheta_n)}(t) dt\\
  &= \int_{\Real^p}  |u_k(\by , \hat \btheta_n + \lambda_t
    t/\sqrt{n})| |t| f_{\sqrt{n}(\Rtheta-\hat \btheta_n)}(t) dt \quad
  \text{ where $0\le\lambda_t\le1$} \\
  &= \int_{\Real^p} |u_k(\by , \hat \btheta_n + \lambda_t t/\sqrt{n})| |t| \phi(t; 0,
 I^{-1}(\btheta_0)) dt\\ & \quad + \int_{\Real^p} |u_k(\by , \hat \btheta_n + \lambda_t t/\sqrt{n})|
 |t| \left[f_{\sqrt{n}(\Rtheta-\hat \btheta_n)}(t)-\phi(t; 0, I^{-1}(\btheta_0))\right] dt
\end{align*}
For the first integral, since $u_k(\by, \hat \btheta_n + \lambda_t t/\sqrt{n}) \cip 0$ as $n\to
\infty$ and the integrand is dominated, by dominated coverage theorem, it goes to 0 in probability.
For the second integral, since $u_k$ is bounded and $\int_{\Real^p} |t| f_{\sqrt{n} (\Rtheta-\hat
\btheta)}(t) dt \cip \int_{\Real^p} |t| \phi(t; 0, I^{-1}(\btheta_0)) dt$, it also goes to 0 in
probability. Finally, equation \eqref{eq:prop3.3} follows directly from the definition of $R_k^*(A)$
and \eqref{eq:prop3.2}.

The proposition can be immediately relaxed for the case $u_k(\by,\cdot)$ is bounded with some
polynomial in $\btheta$ with probability tending to 1. To do this, we have to replace assumption
(E2) by a similar condition with higher moment.

\end{proof}

\begin{proof}[Proof of Proposition 4.]

First, for $\varepsilon>0$, consider
{
  \allowdisplaybreaks
  \begin{align}
  &P\left\{ \left|
   \sqrt{n}\frac{d_k}{T}\left[\sum_{t=0}^{T } \tilde{w}_k(\btheta_{k,t})
   -\frac{J(\by_k, \hat \btheta_n)}{J(\by,\hat \btheta_n)} \sum_{t=0}^{T } w_k(\btheta_{k,t}) \right]
   \right| > \varepsilon  \Big| \by \right\} \nonumber\\
   =&  P\left\{ \left|
   \frac{d_k}{T}\sum_{t=0}^{T } w_k(\btheta_{k,t})   \sqrt{n} \left[
    \frac{J(\by_k, \btheta_{k,t})}{J(\by,\btheta_{k,t})}
    -\frac{J(\by_k, \hat \btheta_n)}{J(\by,\hat \btheta_n)}
    \right] \right| > \varepsilon  \Big| \by \right\} \nonumber\\
    \le& \frac{1}{\varepsilon}
    E \left[\left|
        \frac{d_k}{T}\sum_{t=0}^{T } w_k(\btheta_{k,t})   \sqrt{n} \left[
        \frac{J(\by_k, \btheta_{k,t})}{J(\by,\btheta_{k,t})}
        -\frac{J(\by_k, \hat \btheta_n)}{J(\by,\hat \btheta_n)}
        \right] \right| \Big| \by \right]\nonumber\\
    \le& \frac{d_k}{\varepsilon}
    \frac{1}{T}\sum_{t=0}^{T } E \left[w_k(\btheta_{k,t})   \sqrt{n} \left|
        \frac{J(\by_k, \btheta_{k,t})}{J(\by,\btheta_{k,t})}
        -\frac{J(\by_k, \hat \btheta_n)}{J(\by,\hat \btheta_n)}  \right|  \Big| \by  \right]
  \nonumber\\
    =& \frac{1}{\varepsilon} E \left[ 1\{\Rtheta \in A\}  \sqrt{n} \left|
    \frac{J(\by_k,  \Rtheta)}{J(\by, \Rtheta)}
    -\frac{J(\by_k, \hat \btheta_n)}{J(\by,\hat \btheta_n)} \right|  \Big| \by   \right] \nonumber \\
    \le&
    \frac{1}{\varepsilon} E \left[ \sqrt{n} \left|
    \frac{J(\by_k,  \Rtheta)}{J(\by, \Rtheta)}
    -\frac{J(\by_k, \hat \btheta_n)}{J(\by,\hat \btheta_n)} \right|  \Big| \by   \right]
    \cip 0, \label{eq:prop4.eq1}
  \end{align}

}
as $n\to \infty$, by Proposition 3.
Similarly,
\begin{align}
P\left\{ \left|
 \sqrt{n}\frac{d_k}{T}\left[\sum_{t=0}^{T } 1\{\btheta_{k,t} \in A\} \tilde{w}_k(\btheta_{k,t})
 -\frac{J(\by_k, \hat \btheta_n)}{J(\by,\hat \btheta_n)} \sum_{t=0}^{T } 1\{\btheta_{k,t} \in A\} w_k
 (\btheta_{k,t}) \right]
 \right| > \varepsilon  \Big| \by \right\} \cip 0.\label{eq:prop4.eq2}
\end{align}

Recall that  \[
\tilde R_k(A) = \frac{T^{-1}\sum_{t=0}^{T }1\{\btheta_{k,t} \in A\} \tilde{w}_k(\btheta^{
(t)}_k)}{T^{-1}\sum_{t=0}^{T } \tilde{w}_k(\btheta_{k,t})}.
\]
Equation \eqref{eq:prop4.eq1} and \eqref{eq:prop4.eq2} imply that the numerator and denominator of
$\tilde R_k(A)$ could well approximate, up to a constant, the numerator and denominator of
$\hat R_k (A)$ in~\eqref{eq:importance.finite}. By properties of convergence in probabilities,
we have for large enough $T$ and any $\varepsilon$, as $n \to \infty$,
  \begin{align*}
   P \left[\sqrt{n}\left|\tilde R_k(A)- \hat R_k(A)\right| > \varepsilon \Big| \by \right] \cip 0.
  \end{align*}
Secondly, by Proposition 1, we have, $\sqrt{T}(\hat R_k(A) - R(A))$ given $\by$ is stochastically
bounded. Finally,
  \begin{align*}
    & P\left[ \sqrt{n}\left|\tilde R_k(A) - R(A)|\by \right| > \varepsilon \Big| \by \right] \\
    \le&
    P\left[\sqrt{n}\left|\tilde R_k(A)- \hat R_k(A)\right| + \sqrt{n}\left|\hat R_k(A) - R(A)|\by\right| > \varepsilon \Big| \by \right] \\
    =& P\left[ \sqrt{n}\left|\tilde R_k(A)-
    \hat R_k(A)\right| + \sqrt{\frac{n}{T}} \sqrt{T} \left|\hat R_k(A) - R(A)|\by\right| >
    \varepsilon \Big| \by \right]\cip 0,
  \end{align*}
  $T \to \infty$, $n \to \infty$, $n/T \to 0$.
\end{proof}


\bibliographystyle{rss}
\bibliography{Bibliography}

\end{document}